\documentclass[12pt,preprint]{aastex}

\begin{document}

\title{QSOs Associated with Messier 82}

\author{E. Margaret Burbidge\altaffilmark{1}, Geoffrey
Burbidge\altaffilmark{1}, Halton C. Arp\altaffilmark{2}, and Stefano Zibetti\altaffilmark{2}}

\altaffiltext{1}{Center for Astrophysics \& Space
Sciences, University of California, San Diego}
\altaffiltext{2}{Max-Planck-Institut f\"{u}r
Astrophysik, Garching}

\begin{abstract}

The starburst / AGN galaxy M82 was studied by Dahlem, Weaver and
Heckman using X-ray data from ROSAT and ASCA, as part of their
X-ray survey of edge-on starburst galaxies.  They found seventeen
unresolved hard-X-ray sources around M82, in addition to its
strong nuclear source, and other X-rays within the main body of
M82.  We have measured optical point sources at these positions,
and have obtained redshifts of six candidates at the Keck I 10-m
telescope, using the low-resolution imaging spectrograph (LRIS).
All six are highly compact optical and X-ray objects
with redshifts ranging
from 0.111 to 1.086. They all show emission lines.  The three with
the highest redshifts are clearly QSOs.  The others with lower
redshifts may either be QSOs or compact emission-line galaxies. In
addition to these six there are nine QSOs lying very close to M82
which were discovered many years ago. There is no difference
between optical spectra of these latter QSOs, only two of which
are known to be X-ray sources, and the X-ray emitting QSOs. The
redshifts of all fifteen range between 0.111 and 2.05. The large
number of QSOs and their apparent association with ejected matter
from M82 suggest that they are physically associated with the
galaxy, and have large intrinsic redshift components.  If this is
correct, the absolute magnitudes lie in the range $ -8 < {M_v} <
-10$.  Also we speculate that the luminous variable X-ray source
which has been detected by Chandra in the main body of M82 some
9$^{\prime\prime}$ from the center is another QSO in the process
of ejection from the nucleus, and propose some observational tests of
this hypothesis.

\end{abstract}
\keywords{galaxies: individual (M82) --- galaxies: starburst --- quasars: emission lines --- X-rays: galaxies}

%\keywords{galaxies: individual (M82)
%--- galaxies: starburst --- (galaxies:)  quasars:  emission lines
%--- X-rays: galaxies}

%\vspace{.2in} \noindent
\textbf{1.  INTRODUCTION} \\

\noindent M82 is the nearest active (starburst) galaxy to the
Milky Way. Early optical evidence for an outburst in its center
was found by Lynds and Sandage (1963), who measured outward
velocities in hydrogen, ionized nitrogen and oxygen in the
filamentary structures both sides of this nearly edge-on galaxy.
They deduced that the outward-flowing gas was the result of an
explosive event in the nucleus some 1.5 x 10$^6$ years ago.  More
recently, it has been shown that the outflow observed by Lynds and
Sandage, and by later workers, is related to the tremendous
star-forming activity in the central regions of the galaxy
(Telesco \& Harper 1980).  As well as the optical evidence using
H$\alpha$ (cf. Burbidge et al. 1964; Devine \& Bally 1999), a
large number of compact radio sources identified as supernova
remnants and compact H $\alpha$ regions have been seen (Strickland
et al. 1997; Pedlar et al. 1999, Griffiths et al. 2000).  It
appears that large amounts of molecular gas are
also present (Walter et al. 2002). \\

The first QSOs close to M82 were found serendipitously by Arthur
Hoag, who detected three faint stellar objects with emission lines
on an objective-prism/grating plate taken on the field around M82.
Spectra of these were obtained at Lick Observatory with the
Wampler-Robinson image-dissector scanner on the 3-m telescope, and
revealed all three as QSOs with very similar redshifts, z = 2.048,
2.054, 2.040 respectively (Burbidge et al. 1980).  The similarity
of the redshifts first suggested that perhaps they were associated
with a distant cluster of galaxies and QSOs which accidentally
lies very close to the line of sight to the center of M82.
However, doubt was immediately cast on this suggestion when one of
us (Arp 1983) found a fourth QSO in the same small area with a
very different redshift (z = 0.85, see Table 2). \\

Following these studies a detailed investigation of a field SE of
M82 close to the field in which the first four QSOs had been found
was undertaken by Afanas$^,$ev et al. (1990), Boller (1988), and
Boller et al. (1989).  In these studies the limit that was set was
m$_b$ = 22.0.  Five more QSOs in this area were discovered from
these observations with the Russian 6-m telescope.  This gave a
total of nine QSOs lying in a cone extending about 10$^\prime$ SE
of M82. This was the situation prior to the publication of an
important paper by Dahlem et al. (1998). \\

\noindent
\textbf{2. UNRESOLVED X-RAY SOURCES AROUND M82} \\

\noindent An X-ray survey of edge-on starburst galaxies was
carried out by Dahlem et al. (1998), using data from ROSAT (both
position-sensitive proportional counter and high-resolution
imaging data) and ASCA.  A map by Dahlem et al. of the X-ray data
centered on M82, and covering a field 35$^\prime$ by 35$^\prime$,
shows strong emission centered on the nucleus and extending over
an elliptical area in the directions NNW to SSE where the optical
filaments of Lynds and Sandage are located on both sides of M82,
and where ASCA data has pinpointed many sources due to the
star-formation/supernova activity within M82 (cf. Griffiths et al.
2000). In addition, Dahlem et al. mapped 17 unresolved sources
outside, but close to, M82 (see Fig. 13 of Dahlem et al.). They
noted the similarities of the thermal temperatures of these
sources to those of components within the body of M82, and pointed
out that this suggests ``that at least some of the unresolved
sources are associated with M82". They also pointed out that it
was statistically very unlikely to have so many background point
sources so close. \\

Comparing positions of point sources in the field (Strickland et
al. 1997) and a few known optical positions, one of us (H. C.
Arp), was able to apply zero point corrections of (-2s,
-18$^{\prime\prime}$) to the Dahlem positions and thereby make
optical identifications of most of those point X-ray sources (see
Fig.  1).  Where more than one optical object was in the error
circle, quick-look spectroscopic data enabled us to eliminate a
few objects that are faint foreground stars.  Following this
procedure, we concluded that ten objects are strong candidate
X-ray sources and they were observed by E. M. Burbidge with the
low resolution spectrograph (LRIS) on the Keck I telescope on
March 20, 2002, through considerable and variable cirrus cloud.
Of the 10 objects observed, three turned out to contain objects
which had no features from which to determine redshifts. One was a
blue galaxy, not measured, and the remaining 6 furnished emission
line redshifts.   The average difference between the X-ray
position and optical candidate for the six new emission-line
objects was 19 arc sec.  In the three cases where featureless
continua were observed there were no evidences from absorption
lines or colors that they were stars.
Therefore they would remain as candidates for BL Lac type spectra associated with M82.\\

\noindent
\textbf{3. RESULTS} \\

\noindent The six Dahlem et al. emission-line sources that we
observed turned out to be low-to-medium redshift QSOs or QSO-like
objects (Table 1 and Fig. 2, which shows the spectra). \\

The three with the highest redshift objects show typical QSO
spectra.  For the lower redshift objects, only the red end of the
spectra showed clearly measurable features.    Details of the spectra are as follows: \\

\begin{tabular}{ll}
Dahlem 7: & (MgII 2798, [OII] 3727, [NeIII] 3869, H$\gamma$,
[HeII] 4686, \\
& H$\beta$, [OIII] 4959, 5007)\\
Dahlem 12: & (MgII 2798, H$\gamma$, H$\beta$, [OIII] 4959, 5007)\\
Dahlem 17: & (SiIII] + CIII] blend 1900, CII 2326, [NeIV] + FeIII
blend 2423, MgII 2798)\\
Dahlem 3: & ([OII] 3727, H$\alpha$, [SII] blend 6725)\\
Dahlem 4: & (H$\alpha$, [SII] blend 6725)\\
Dahlem 9: & ([NII] 6548, H$\alpha$, [NII] 6583, [SII] 6717, 6731)
\end{tabular} \\

\noindent We believe that the three low redshift objects are QSOs,
but we cannot exclude the possibility that they are compact
emission-line galaxies.

All of the QSOs and QSO-like objects now known to lie within 30
arc minutes of M82 are listed in Tables 1 and 2. There are 15 and
they do not appear to be randomly distributed. In Figure 1 we plot
all 15 around an image of M82 which was taken in its H$\alpha$
emission line.  We note the following:

\begin{enumerate}
\item The 9 QSOs SSE of M82 have on the average, higher redshifts than  the 6 N of
M82.

\item There is a comparatively close pair, BOL75 and M82 No. 69,
which has a separation of only 39$^{\prime\prime}$ (630 pc at the
distance of M82), and a triple system Hoag 1, Hoag 2 and Hoag 3
with separations of 118$^{\prime\prime}$ (1-2),
62$^{\prime\prime}$ (2-3) and 195$^{\prime\prime}$ (3-1) and
similar redshifts.

\item All 15 objects with redshifts, of which 12 are certainly QSOs,
and 3 are emission line objects which may also be QSOs, lie within
15$^{\prime}$ of the center of the galaxy.  This gives
conservatively a density of at least 61 QSOs per square degree. Of
the 12 QSOs, 6 are brighter than 19$^{m}$.5, 8 are brighter than
20$^{m}$, and 10 are brighter than 20$^{m}$.5. These correspond to
densities of 30, 41 and 51 per square degree respectively. Such
densities are to be compared with those obtained in QSO surveys by
Kilkenny et al. (1997) and Boyle et al. (2000) which give
respectively 10 per square degree to 20$^{m}$, and 25 per square
degree for 18.25 $<$ b$_j$ $<$ 20.85 from the 2dF survey with the
Anglo Australian Telescope. While there are small uncertainties
associated with the magnitude calibrations, and the total numbers
are small, they do mean that there is a significant over-density
of QSOs in the magnitude range down to 20$^{m}$ - 20$^{m}$.5 near
to M82 compared with those in the general field.

\item The distribution of these QSOs is far from uniform.  No
QSOs have been detected SW of M82 down to the magnitudes of the QSOs SE of M82.
There are candidates among the
X-ray sources listed by Dahlem et al.  immediately SE of M82, but the
distribution of unresolved X-ray sources as shown in their Fig. 13
and our Figure 1 quite definitely shows the concentrations NW to
N, and S to SE.

\item As can be seen in Fig. 1, there is a long H$\alpha$ filament,
apparently ejected along the minor axis NNW from M82 (Devine and
Bally 1999).  Just in the vicinity of the ``cap" at about $\theta$
= 11 arc min we find the strong X-ray emitting object with z =
0.111. This suggests that the point X-ray sources have been
entrained, moving along the path followed by the z = 0.68 QSO
through the z = 0.11 object and on to the z = 0.63 QSO and the z =
1.09 QSO. There are many point X-ray sources found from the high
resolution ROSAT (HRI) map which are apparently emerging from the
body of M82.

\item The reddening and absorption in active galaxies like M82 is
known to be large.  In this connection it is noticeable that in
Table 1 the sources on the NNW side of M82 appear generally bluer
and brighter than the sources on the SW side.  This might imply
that these QSOs are on the near side of M82. If this is correct it
would imply that the 11 ``background" QSOs suffer absorption and
hence their density would be expected to be lower than it is in
the central region.  This means that the density of M82 quasars
should be compared to a background density QSOs which are about a
magnitude brighter, and thus less frequent.

Finally we turn briefly to the numerical values of the redshifts
in Tables 1 and 2.
\end{enumerate}

In a number of investigations of QSOs and active galaxies it has
been shown that QSO redshifts tend to peak about certain values: z
= 0.061, 0.30, 0.60, 0.96, 1.41, 1.96, etc. and that these peaks
are periodic with $\triangle$ log (1 + z) = 0.089 (Burbidge and
Napier 2001 and earlier references given there).  In the most
general case the observed redshift z$_0$ is given by \\

\begin{center}
z$_0$ = [(1 + z$_c$)(1 + z$_d$)(1 + z$_i$)] - 1 \\
\end{center}

\noindent where z$_c$, z$_d$ and z$_i$ are the cosmological,
Doppler and intrinsic redshift components.  Where the sample is
made up of QSOs associated with bright low-redshift galaxies, we
know that z$_c$ in each case is very small.  Also, z$_d$ is a
measure of the projected speed with which the QSOs are ejected
from the galaxies. It turns out that for the X-ray emitting QSOs
z$_d$ $\approx |0.04|$ (Burbidge and Napier 2001). Thus, for the
QSOs associated with M82 z$_0 \simeq z_i$. We might expect,
therefore, that many of the values of z$_0$ in Table 1 might lie
close to the peak intrinsic values given above. This is clearly
the case. There are two close to 0.60 (0.626, 0.675), three very
close to 0.96 (0.93, 0.96, 1.01), and three very close to 1.96
(2.040, 2.048, 2.054). We conclude that the high surface density
of these QSOs together with their remarkable distribution about
M82, and finally the concentration of the redshifts about
previously determined intrinsic values, all suggest that the QSOs
have been ejected from M82.

M82 is the nearest active galaxy around which many QSOs have been
detected.  If they are physically associated with M82, since the
distance modulus is 27.7, they are all intrinsically faint, with
absolute magnitudes Mv in the range -8 to -10.

All of the other active galaxies around which QSOs have been
found, such as NGC 4258 (Burbidge 1995), NGC 2639 (Burbidge 1997),
NGC 1068 (Burbidge 1999), Arp 220 (Arp, Burbidge 2001) or NGC 3628
(Arp, Burbidge 2002) are much further away than M82.

The very different distances of those galaxies means that the QSOs
found near to them with apparent magnitudes 18$^m$ - 20$^m$ are
intrinsically much brighter than the QSOs around M82 described
here.  For example, for the case of Arp 220, which lies at a
distance of about 90Mpc (Ho = 60 Km sec$^{-1}$ Mpc$^{-1}$) the
QSOs about 10$^{\prime}$ - 20$^{\prime}$ from its center have Mv =
-14.5 - 15.5. Thus if Arp 220 has faint QSOs near it at the same
projected linear distances as those near to M82 and similar to
those QSOs, they will lie $\lesssim$ 1$^\prime$ - 2$^\prime$ from
the nucleus, and will have apparent magnitudes $\sim$ 24$^m$ -
25$^m$. Thus while they will contribute to the central luminosity
of Arp 220, they will be so faint that they will not be detectable
as individual optical sources.

However, if M82 has QSOs as bright as those found around Arp 220
associated with it, the equivalent linear distances from the
center of M82 will be ten times as far away from the center of
that galaxy, i.e. distances of 2 - 4 degrees.  QSOs at such large
angular distances will not easily be identified as being
associated with M82.  They would have apparent magnitudes 15 or
brighter, i.e. some of the brightest QSOs might have originated
from such local galaxies.  (In fact, brighter QSOs of this angular
distance range from M82 have been pointed out by Arp and Russell
(2001, Fig. 6).  Thus it may be the case that active galaxies can
give rise to many QSOs with a wide range of luminosities both in
optical flux and in X-rays.  Their detection will then depend on
how far from the nucleus they have travelled, and how bright they
are.

\noindent
\textbf{4.  POSSIBLE EVIDENCE FOR QSOs INSIDE M82} \\

\noindent Studies over many years of the central regions of
comparatively nearby starburst galaxies, and galaxies with active
galactic nuclei, have shown that they often contain compact
sources outside the centers which have been detected both as
discrete radio sources and discrete compact X-ray sources.  Some
of these sources are known to be highly variable.  In several high
resolution X-ray studies using ASCA and Chandra, a number of
luminous X-ray sources have been detected.  Apart from those
resulting from supernovae, elsewhere we have made the general
argument that sources of this kind are likely to be local QSOs in
the process of ejection from the nucleus of the active galaxy
(Burbidge, Burbidge \& Arp 2003). In particular, a luminous
variable X-ray source CXO M82J095550.2 + 694047 has been detected
by ASCA and Chandra some 9$^{\prime\prime}$ from the center of M82
and it shows large time variability. Its luminosity has varied
between about 10$^{40}$ and 10$^{41}$ erg/sec in the 0.2-10 kev
energy range (Matsumoto et al. 2001; Kaaret et al. 2000).

We believe that this may be a newly born QSO being ejected from the
nucleus of M82.
There are several tests of this hypothesis which are possible.  If
the hypothesis is correct, we would expect that if any
spectroscopic features could be detected in X-ray, radio, or
optical wavelengths, they would show anomalous redshifts as do the
QSOs outside the galaxy.  Apparently such an object has recently
been detected in the galaxy NGC 4868 (Foschini et al. 2002a, b).
\\

Secondly, based on X-ray emitting QSOs, we estimate that these
objects will be ejected at speeds of 0.1 c.  Thus, for a galaxy as
close as M82, it may be possible to measure proper motions. For
ejection speeds in the range 10000 km sec$^{-1}$ to 30000 km
sec$^{-1}$, we might detect motions of about 6-18 milliarc sec/yr.
Since the sources are variable in flux, it may be hard to
disentangle variability from the movement of the center
of gravity of the source.    However, a proper motion search could  provide a
practical test within 5--10 years. \\

We wish to thank Vesa Junkkarinen for his help with obtaining the
data at Keck Observatory.

\clearpage
\begin{center}
\tiny{
\begin{tabular}{lllllllrrcrrr}
\multicolumn{13}{c}{TABLE 1} \\
\multicolumn{13}{c}{QSOS and Other Objects Near M82} \\

\hline\hline \\
& & & & & & & & & &  {Ang. Dist.} & X-ray & $\triangle$ Position \\
OBJECT & \multicolumn{3}{c}{$\alpha$(2000)} &
\multicolumn{3}{c}{$\delta$(2000)} & mag. & color & redshift &
from M82 & cts/ks & arc sec \\
& & & & & & & & & & arc minutes & & \\
\hline & & & & & & & & & & & & \\
& h & m & s & $^\circ$ & $^\prime$ & $^{\prime\prime}$ & E & O-E & z &  $\theta$ &
C & $\triangle$ r \\
\hline & & & & & & & & & & & & \\
Dahlem 3 & 9 & 53 & 18.9 & +69 & 45 & 44 & 18.7 & 1.1 & 0.190 & 14.2 &  6 & 6 \\
Dahlem 4 & 9 & 53 & 35.3 & +69 & 47 & 51 & 18.6 & .4 & 0.221 & 13.8 & 9 & 25 \\
Dahlem 7 & 9 & 55 & 18.8 & +69 & 47 & 00 & 19.8 & $\sim$.0 & 0.675 & 6.9 & 9.7 & 31 \\
Dahlem 9 & 9 & 55 & 32.7 & +69 & 51 & 34 & 18.9 & .4 & 0.111 & 10.9 & 18.4 & 27 \\
Dahlem 12 & 9 & 56 & 41.0 & +69 & 52 & 01 & 18.9 & .2 & 0.626 & 12.0 &  2.6 & 7.4 \\

\vspace{.2in}

Dahlem 17 & 9 & 57 & 16.4 & +69 & 54 & 25 & 17.2 & .1 & 1.086 & 5.4 & 2.5 & 17 \\
Dahlem 6 & 9 & 54 & 34.1 & +69 & 48 & 37 & $>$19.2 & blue & continuum &  & 1.2 & 20 \\
Dahlem 8 & 9 & 55 & 16.5 & +69 & 51 & 15 & 18.2 & .9 & continuum & & 10 & 19 \\
Dahlem 10 & 9 & 55 & 36.4 & +69 & 55 & 07 & 17.3 & -.2 & blue gal & & 12 & --- \\
Dahlem 15 & 9 & 56 & 56.4 & +69 & 34 & 05 & 18.9 & .4 & continuum & & 3.9 & 24 \\
Dahlem 14 & & & & & & & & & QSO Hoag 1 & & & \\
Dahlem 1 & & & & & & & & & no data & & & \\
Dahlem 2 & & & & & & & & & no data & & & \\
Dahlem 5 & & & & & & & & & no data & & & \\
Dahlem 11 & & & & & & & & & no ident & & & \\
Dahlem 13 & & & & & & & & & no ident & & & \\
Dahlem 16 & & & & & & & & & QSO NGC3031 U4 & & & \\
\end{tabular}}
\end{center}

\clearpage

%\vspace{.15in}
\noindent The upper part of Table 1 lists the 6
Dahlem objects that we observed, the 2000 coordinates from Dahlem
et al., the magnitudes and colors of our identified optical
objects, our measured redshifts, their angular distances from the
center of M82 in arc minutes, their X-ray counts per kilosecond,
and the angular distance in arc seconds of the optical object from
the X-ray position.  The lower part gives the same data for 4 more
Dahlem objects which we observed but did not obtain redshifts,
Dahlem 14 and 16 which were already known to be QSOs with known
redshifts, three candidate identifications for which we obtained
no data, and two sources with no optical candidate
identification. \\

\clearpage

\begin{center}
\tiny{
\begin{tabular}{lllllllrrcccc}
\multicolumn{13}{c}{TABLE 2} \\
\multicolumn{13}{c}{QSOS Identified Earlier Near M82} \\

\hline\hline \\
& & & & & & & & & &  {Ang. Dist.} & & $\triangle$ Position \\
OBJECT & \multicolumn{3}{c}{$\alpha$(2000)} &
\multicolumn{3}{c}{$\delta$(2000)} & mag. & color & z &
from M82 & C & arc sec \\
& & & & & & & & & & arc minutes & & \\
\hline & & & & & & & & & & & & \\
& h & m & s & $^\circ$ & $^\prime$ & $^{\prime\prime}$ & V & B-V &
z & $\theta$ & C & $\triangle$ r \\
\hline & & & & & & & & & & & & \\
M82 No. 95 & 9 & 56 & 41.9 & +69 & 39 & 24 & 19.44 & .36 & 1.01 & 4.55  & --- & --- \\
Hoag 1 & 9 & 56 & 58.2 & +69 & 38 & 37 & 19.26 & .84 & 2.048 & 6.15 & 2.8 & 11 \\
Hoag 2 & 9 & 57 & 19.8 & +69 & 58 & 01 & 19.79 & .76 & 2.054 & 8.11 & --- & --- \\
NGC3031 U4 & 9 & 57 & 20.2 & +69 & 35 & 37 & 20.1 & .70 & .85 & 9.23 &  3.9 & 17 \\
Hoag 3 & 9 & 57 & 23.5 & +69 & 36 & 13 & 20.51 & .69 & 2.040 & 9.17 & --- & --- \\
Bol 105 & 9 & 57 & 36.9 & +69 & 38 & 12 & 21.4 & --- & 2.24 & 9.45 & --- & --- \\
M82 No. 69 & 9 & 57 & 49.5 & +69 & 37 & 12 & 19.38 & .70 & 0.93 & 10.81 & --- & --- \\
M82 No. 22 & 9 & 57 & 55.8 & +69 & 32 & 56 & 19.04 & 1.31 & 0.96 & 13.31 & --- & --- \\
Bol 75 & 9 & 57 & 57.0 & +69 & 37 & 08 & 22 & --- & 0.74 & 11.45 & ---  & --- \\
\end{tabular}}
\end{center}

\clearpage
\begin{center}
\textbf{REFERENCES} \\
\end{center}

\begin{description}
\item Afanas'ev, V. L., Vlasyuk, V. V., Dodonov, S. N., Lorenz, H. \& Terebizh, V. Y., 1990,
Astrof. Issled:Izvestiya Spetsialnoi Astrofiz. Obs., 32, 31.

\item Arp, H. C., 1983, ApJ., 271, 479.

\item Arp, H. C., 1999, ApJ., 525, 594.

\item Arp, H. C., Burbidge, E. M., Chu, Y. and Zhu, X., 2001,
ApJ., 553, L11.

\item Arp, H. \& Russell, D., 2001, ApJ., 549, 802.

\item Boller, T., 1988, thesis, ZIAP AS GDR.

\item Boller, T., Lorenz, H., Afanas'ev, V. L., Dodonov, S. N. \&
Terebizh, V. Y., 1989, Astr. Nachr., 310, 3, 187.

\item Boyle, B. J., Shanks, T., Croom, S. M., Smith, R. J., Miller, L., Loaring, L. \& Heymans, C., 2000,
MNRAS, 317, 1014.

\item Burbidge, E. M., 1995, A\&A, 298, L1.

\item Burbidge, E. M., 1997, ApJ., 477, L13.

\item Burbidge, E. M., 1999, ApJ., 511, L9.

\item Burbidge, E. M., Burbidge, G. R. \& Rubin, V., 1964, ApJ., 140,  942.

\item Burbidge, E. M., Junkkarinen, V. T., Koski, A. T., Smith, H.
E. \& Hoag, A. A., 1980, ApJL, 242, L55.

\item Burbidge, G. \& Napier, W. M., 2001, AJ, 121, 21.

\item Burbidge, G. , Burbidge, E. M., \& Arp, H. C., 2003, A\&A Letters, in press.

\item Dahlem, M., Weaver, K. A., \& Heckman, T. M., 1998, ApJS, 118,  401.

\item Devine, D. \& Bally, J., 1999, ApJ., 510, 197.

\item Foschini, L., Di Cocco, G., Ho, L. C., Bassani, L., Cappi, M., Dadina, M., Gianotti, F.,
Malaguti, G., Panessa, F., Piconcelli, E., Stephen, J. B.,
Trifoglio, M., 2002a, arXiv:  astro-ph/0206418

\item Foschini, L., Di Cocco, G., Ho, L. C., Bassani, L., Cappi, M., Dadina, M., Gianotti, F.,
Malaguti, G., Panessa, F., Piconcelli, E., Stephen, J. B.,
Trifoglio, M., 2002b, arXiv:  astro-ph/0209298

\item Griffiths, R. E., Ptak, A., Feigelson, E. D. et al., 2000,
Science, 290, 1325.

\item Kaaret, P., Prestwich, A., Zezas, A. et al., 2000, MNRAS, 000, 1.

\item Karlsson, K. G., 1971, A\&A, 13, 333.

\item Karlsson, K. G., 1990, A\&A, 239, 50.

\item Kilkenny, D., O'Donoghue, D., Koen, C., Stobie, R. S. \& Chen, A., 1997, MNRAS, 287, 867.

\item Kronberg, P., Biermann, P. \& Schwab, F. R., 1981, ApJ., 246,  751.

\item Lynds, C. R. \& Sandage, A. R. 1963, ApJ, 137, 1005.

\item Matsumoto, H., Tsuru, T. G., Koyama, K., Awaki, H.,
Canizares, C. R., Kawai, N., Matsushita, S. \& Kawabe, R., 2001,
ApJ, 547, L25.

\item Pedlar, A., Muxlow, T., Garrett, M., Diamond, P., Wills, K.
A., Wilkinson, P. \& Alef, W., 1999, MNRAS, 307, 761.

\item Strickland, D., Ponman, T. J. \& Stevens, I. R., 1997, A\&A,
320, 378.

\item Telesco, C. \& Harper, D., 1980, ApJ., 235, 392.

\item Walter, F., Weiss, A. \& Scoville, N., 2002, ApJL, 580, L21.
\end{description}

\newpage
\noindent Fig. 1:  The plot shows all of the QSOs and QSO-like
objects within an area of 0.25 deg.$^2$ centered on M82 (from
Tables 1 and 2). The inserted image is M82 in the light of H
$\alpha$ emission with the filament going NW and ending on the H
$\alpha$ ``cap".  (Image from Devine and Bally 1999.) \\

\noindent Fig. 2:  Point-source X-ray objects with emission-line
redshifts NW of M82 observed with the Low-Resolution Imaging
Spectrometer (LRIS) on the Keck I 10-m telescope.  Our setups used
the 400/8500 grating, 1.86\AA/pixel, on the red side and the
400/3400 grism, 1.09\AA/pixel, on the blue side. From top left:
Dahlem 7, z=0.675; Dahlem 12, z =0.626; Dahlem 17, z =1.086. From
top right:  Dahlem 3, z=0.190; Dahlem 4, z=0.221; Dahlem 9,
z=0.111.

\noindent Fig. 3:  The points are catalogued X-ray sources from
the ROSAT high resolution (HRI) imager.  Circled objects have
redshifts from Tables 1 and 2.  The inner point sources appear to
be part of the gas ejected from the nucleus of M82.  A principal
outflow of molecular gas has been mapped in just the direction of
the ESE QSOs at p.a. = 110 deg. (Walter et al. 2002, Fig. 1).

%%%UCP%%%
\newpage
\plotone{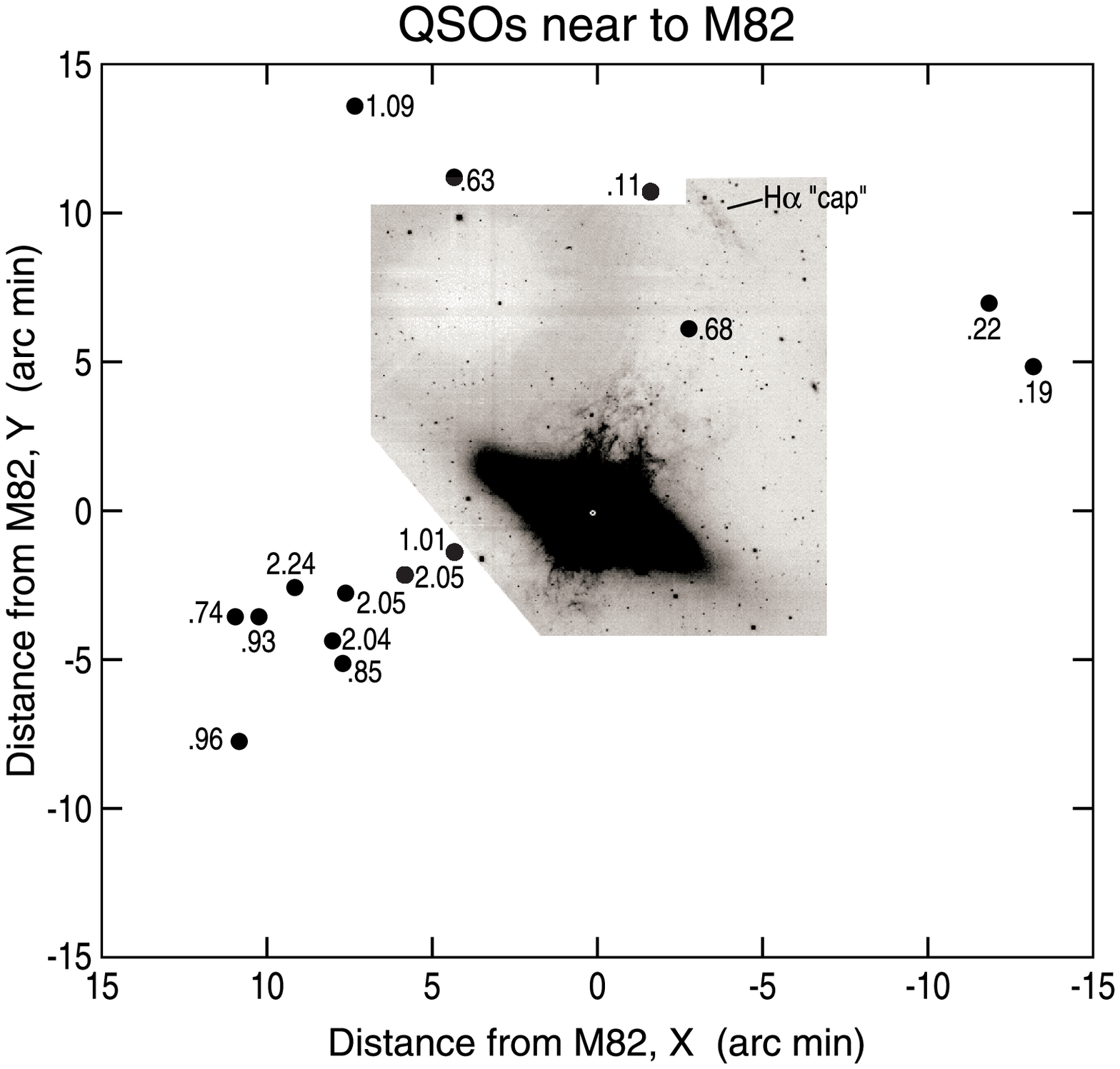}

\newpage
\plotone{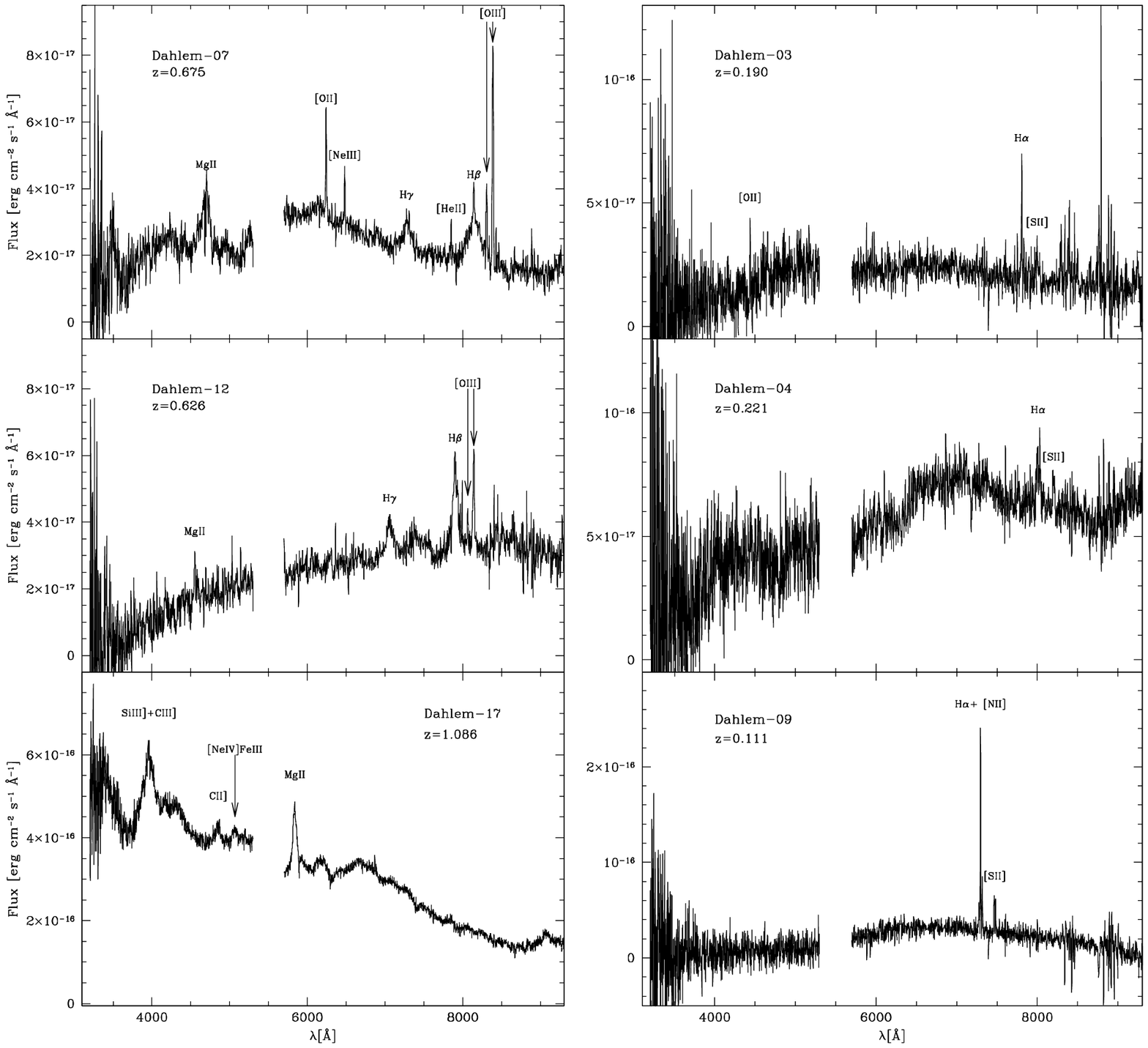}

\newpage
\plotone{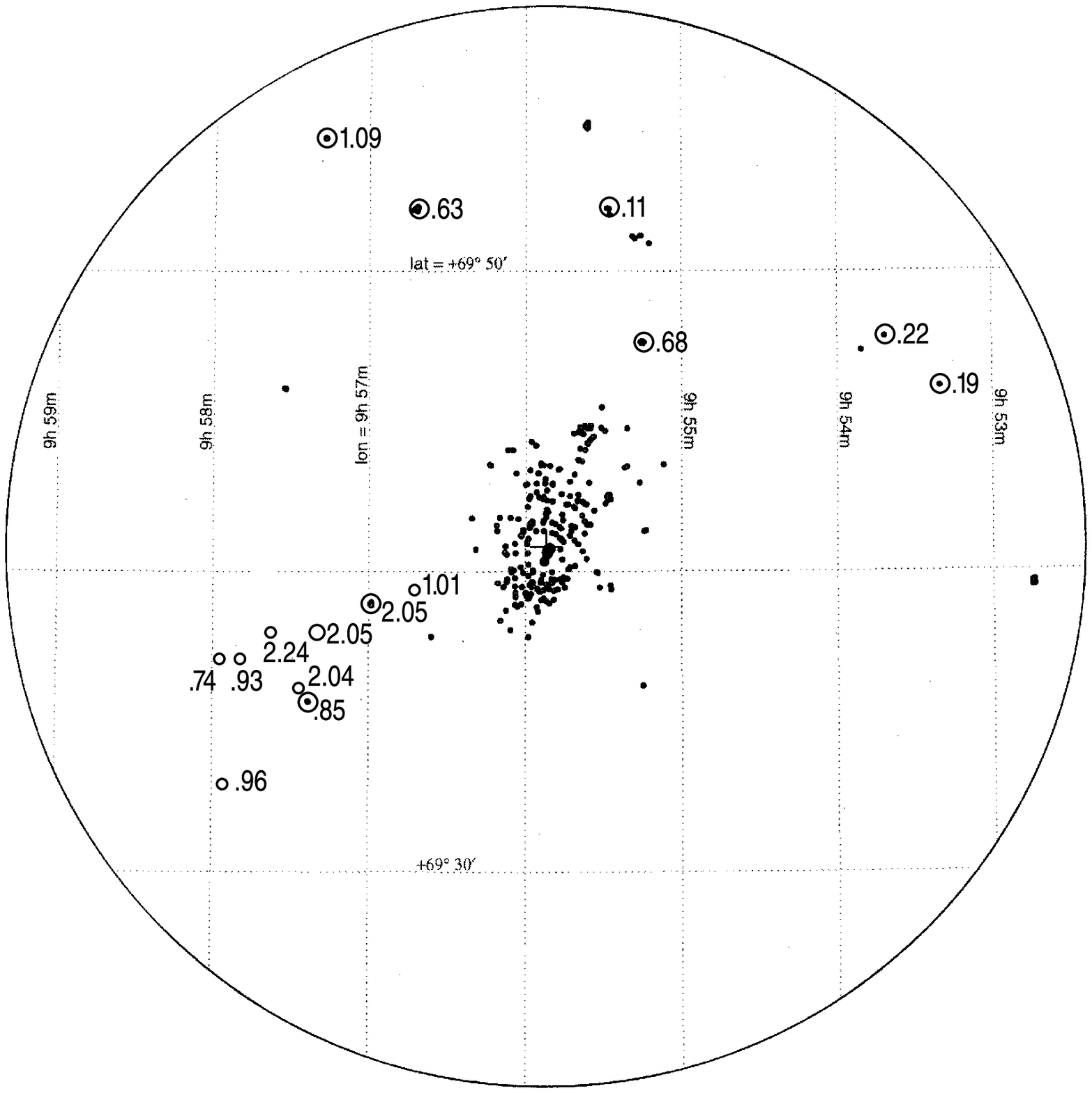}

\end{document}